\documentclass[sigconf, screen, authorversion]{acmart}

\copyrightyear{2023}
\acmYear{2023}
\setcopyright{rightsretained}
\acmConference[CUI '23]{ACM conference on Conversational User Interfaces}{July 19--21, 2023}{Eindhoven, Netherlands}
\acmBooktitle{ACM conference on Conversational User Interfaces (CUI '23), July 19--21, 2023, Eindhoven, Netherlands}\acmDOI{10.1145/3571884.3603760}
\acmISBN{979-8-4007-0014-9/23/07}

\begin{document}

\title{Who are CUIs Really For? Representation and Accessibility in the Conversational User Interface Literature}

\author{William Seymour}
\email{william.1.seymour@kcl.ac.uk}
\orcid{0000-0002-0256-6740}
\affiliation{%
  \institution{King's College London}
  \streetaddress{Bush House, 30 Aldwych}
  \city{London}
  \country{UK}
  \postcode{WC2B 4BG}
}

\author{Xiao Zhan}
\email{xiao.zhan@kcl.ac.uk}
\orcid{0000-0003-1755-0976}
\affiliation{%
  \institution{King's College London}
  \streetaddress{Bush House, 30 Aldwych}
  \city{London}
  \country{UK}
  \postcode{WC2B 4BG}
}

\author{Mark Cot\'{e}}
\email{mark.cote@kcl.ac.uk}
\affiliation{%
  \institution{King's College London}
  \streetaddress{Chesham Building, Strand}
  \city{London}
  \country{UK}
  \postcode{WC2R 2LS}
}

\author{Jose Such}
\email{jose.such@kcl.ac.uk}
\affiliation{%
  \institution{King's College London}
  \streetaddress{Bush House, 30 Aldwych}
  \city{London}
  \country{UK}
  \postcode{WC2B 4BG}
}

\renewcommand{\shortauthors}{Seymour and Zhan et al.}

\begin{abstract}
The theme for CUI 2023 is `designing for inclusive conversation', but who are CUIs really designed for? The field has its roots in computer science, which has a long acknowledged diversity problem. Inspired by studies mapping out the diversity of the CHI and voice assistant literature, we set out to investigate how these issues have (or have not) shaped the CUI literature. To do this we reviewed the 46 full-length research papers that have been published at CUI since its inception in 2019. After detailing the eight papers that engage with accessibility, social interaction, and performance of gender, we show that 90\% of papers published at CUI with user studies recruit participants from Europe and North America (or do not specify). To complement existing work in the community towards diversity we discuss the factors that have contributed to the current status quo, and offer some initial suggestions as to how we as a CUI community can continue to improve. We hope that this will form the beginning of a wider discussion at the conference. 
\end{abstract}




\maketitle

\section{Introduction}
Conversational assistants and chatbots continue to increase in popularity, leveraging advancements in machine learning to improve reasoning, natural language processing, and voice recognition. Since their inception, an interdisciplinary research community has grown around voice and conversational technologies, beginning in computer science with the likes of ELIZA~\cite{weizenbaum1966eliza}. It has since expanded to encompass a wide array of disciplines, and this is the fifth year that we have congregated under the CUI banner.

It has long been recognised that both computer science as a discipline and technology as a sector have major diversity problems, which may even be worsening over time\footnote{\url{https://www.wired.com/story/computer-science-graduates-diversity}} through a vicious cycle of underrepresentation and underenrollment that is compounded by structural inequality, bias, and harassment. The multiplicative power of design means that this white male homogeneity extends to the products we use every day, leading to technology that isn't designed for the marginalised majority. It is unsurprising then that in the home we see that the installation and abuse~\cite{10.1145/3411764.3445114} of connected devices has a significant gender component, and that video doorbells reinforce existing racism.\footnote{\url{https://www.vice.com/en/article/qvyvzd/amazons-home-security-company-is-turning-everyone-into-cops}}

But what about research? An analysis of CHI papers from 2016--2020 found that 73\% of CHI studies recruited Western participant samples representing less than 12\% of the world’s population; this reflects the organisations that submit to CHI, as over 80\% of papers used participants from the country of authors' institutions~\cite{10.1145/3411764.3445488}. Inspired by this we have previously investigated these issues in the context of voice assistants and found a similar story. As part of a larger systematic review of ethical concerns with voice assistants, we coded papers for participant demographics and analysed the community's response to issues of equity including accessibility and performance of gender~\cite{seymour2022systematic}. In this provocation we extend this analysis to CUI papers to address an important aspect of the conference theme `designing for inclusive conversation'. To do this we briefly outline the findings from the voice assistant review  before discussing those findings in relation to the papers published at CUI from the inaugural conference in 2019 though to 2022.

\section{The Survey}\label{sec:survey}
We conducted a systematic literature review of the literature on ethical concerns with voice assistants, analysing 117 items published since 2012~\cite{seymour2022systematic}. We adopted a broad view of ethics and related concerns, using the definition of ``what a design object ought to be based on ethical and moral codes'', in contrast with its purpose/function (reason), and visual values/presentation (aesthetics)~\cite{10.1145/3173574.3173578}. The scope of the review included journal articles and conference papers indexed by the ACM Digital Library, IEEE Explore, Web of Science, and DBLP.

The majority of manuscripts that were in-scope for the review, and thus its major findings, were centred around privacy, social interaction, accessibility, performance of gender, and social order. We briefly summarise these here, going into more depth for topics of particular relevance to the provocation:

\subsubsection*{Privacy.} Users' privacy concerns related to voice assistants, including data collection by vendors, sharing of devices between users, and access to devices from those outside the household/home

\subsubsection*{Social Interaction.} Often built on anthropomorphism and the Computers as Social Actors paradigm, these papers explored whether voice assistants are ontologically distinct from people and/or machines, how they conceptualise them, as well as the emotional connections and relationships users form with them

\subsubsection*{Social Order.} How voice assistants affect the dynamics of domestic hierarchies, including parent-child relationships

\subsubsection*{Accessibility}
CUIs are interestingly situated in terms of accessibility. On the one hand, their text/language based conversational nature makes them accessible to the blind and partially sighted~\cite{10.1145/3234695.3236344, 10.1145/3439231.3440603}. On the other hand, for verbal systems those with speech, language, and hearing difficulties are less adequately served~\cite{10.1145/3173574.3174033}, as well as those who have difficulties formulating requests in the manner and speed required~\cite{10.1145/2771839.2771910, 10.1145/3196709.3196772, 10.1145/3173574.3174033, 10.1145/3459990.3465195}. Speech recognition accuracy varies across languages, accents, gender, and other demographics, although this is often framed as a challenge of engineering rather than social inequality~\cite{10.1145/3308560.3317597}.

\subsubsection*{Performance of Gender}
In many products and in popular culture, the default persona for a conversational user interface might be characterised as a white middle-class American woman, with vendors justifying this based on user acceptability with studies suggesting that people perceive women's voices and names as sounding more gentle, kind, and caring than men's~\cite{costa2019ai, schneider2021recommended}. It has been known since the 90's that people apply gender and other stereotypes to computers~\cite{nass1994computers}, and as a research community we have long accepted that we have a duty to push back against harmful preconceptions around gender roles in society. At the same time, attempts to create `genderless' presentations often do not stop users gendering conversational interfaces~\cite{10.1145/3405755.3406123} and can (ironically) erase or ignore those outside the gender binary~\cite{10.1145/3449206}.

\subsection{Additional Analysis}
We also coded papers for the country of residence of their participants (shown in Table~\ref{tab:geomap}) and the specific groups demographics they recruited (shown in Table~\ref{tab:subpops}) where applicable. Of the 85 papers with human participants, \textbf{94\% drew solely or mostly from North America and Europe}. Of the 22 papers that explored the experiences of a particular group of people, 13 were focused on age, 7 on (dis)ability, and 1 each on trans/non binary and Portuguese speakers. Together these represented just over one quarter of the studies that recruited human participants. Within the papers concerned with age there was a diversity of concerns across anthropomorphism (5), accessibility (4), social order (2), and accountability (1). In contrast, all of the papers that recruited based on ability were focused on accessibility concerns. Studies looking at specific groups mainly took qualitative approaches (70\%), suggesting a focus on lived experiences rather than the gathering of a high level overview.

\begin{table}
    \centering
    \caption{Participant country of residence for papers with human participants.}
    \begin{tabular}{lc|lc}
    \toprule
    Country & Papers & Country & Papers \\
    \midrule
    USA & 47 & Netherlands & 2 \\
    UK & 11 & Ireland & 1 \\
    Germany & 10 & South Korea & 1 \\
    Italy & 4 & Spain & 1 \\
    India & 2 & Sweden & 1 \\
    Brazil & 2 & Switzerland & 1 \\
    Canada & 2 && \\
    \bottomrule
    \end{tabular}
    \label{tab:geomap}
\end{table}

\begin{table}
    \centering
    \caption{Number of papers recruiting specific groups, written as reported.}
    \begin{tabular}{r|c}
        \toprule
        Sub-population & \# of Papers \\ \hline
        None & 55 \\
        Children & 10 \\
        Blind and visually impaired & 3 \\
        People with Dysarthria & 2 \\
        Older adults & 2 \\
        Trans and non-binary & 1 \\
        Users with disabilities & 1 \\
        Users with motor impairments & 1 \\
        Portuguese Speakers & 1 \\
        Young Adults (18-36) & 1 \\
        \bottomrule
    \end{tabular}
    \label{tab:subpops}
\end{table}

\section{But it's a CHI Problem, Right?}
We found that 94\% of participants in papers \textit{investigating ethics} were from Europe or North America. That's bad, but those papers were mainly from CHI and CSCW; only 3 of the 117 papers were published at CUI. It's a fair criticism that the timing of the survey ruled out papers from CUI 2022, and that the only captured papers on voice assistants---this ignores the fact that CUI is broader than just voice assistants, with the conference encompassing a host of other conversational interfaces.

To reflect and address this, we conducted a (less rigourous) sift of the 46 full papers that have been published at CUI since its inception. Using the conference proceedings in the ACM Digital Library, we analysed paper abstracts, introductions, and methodologies, comparing them against the themes and demographics from the original review paper. In total, we found eight papers (17\%) exploring topics around accessibility~\cite{10.1145/3405755.3406129, 10.1145/3469595.3469604, 10.1145/3543829.3543839}, social interaction~\cite{10.1145/3342775.3342786, 10.1145/3405755.3406124, 10.1145/3543829.3543835}, and the performance of gender by CUIs~\cite{10.1145/3405755.3406123, 10.1145/3543829.3543836}. By contrast, around half of the provocation papers at CUI 2022 explored questions related to ethics.

For accessibility, CUI papers cover similar themes to the voice assistant review, focusing on enabling children to interact with CUIs on their own terms~\cite{10.1145/3405755.3406129, 10.1145/3469595.3469604}. They also analysed command construction in native and non-native speakers~\cite{10.1145/3543829.3543839}. All accessibility papers used voice as the mode of interaction. The papers on social interaction were also split between categories encountered during the voice assistant literature survey. Two focused on expanding communication with CUIs to encompass speech acts and emotions\cite{10.1145/3405755.3406124, 10.1145/3543829.3543835}, and one on the effect of accented synthetic speech on peoples' responses~\cite{10.1145/3342775.3342786}. Finally, the two papers on performance of gender explored sensitivity when designing gender for voice interfaces~\cite{10.1145/3405755.3406123} and user acceptability of speech with varying gender-linked characteristics~\cite{10.1145/3543829.3543836}.

In terms of participants, \textbf{90\% of CUI papers with user studies recruited participants solely residing in Europe and North America or did not report where participants were recruited from} (38 of 42 papers), with the two papers studying participants from South America sharing common authors (the only studies focusing on locations outside of Europe and North America)~\cite{10.1145/3342775.3342787, 10.1145/3405755.3406128}. The breakdown of participant location by year of the conference is given in Table~\ref{tab:cui_demo}. As a general rule, the geographic location of participants matched that of paper authors. The size of the sample meant that there were no meaningful trends in terms of distribution, apart from a balancing of papers from Europe and the US in more recent years of the conference.

\begin{table}
    \centering
    \caption{Majority participant reigon of residence for CUI papers. For one paper at each of CUI 2020 and CUI 2022 it was not possible to ascertain the participants' region.}
    \begin{tabular}{r|r|r|r|r|r}
        \toprule
        CUI & Europe & N. America & S. America & Other & Multiple \\
        \hline
        2019 & 6 & 2 & 1 & 0 & 0 \\
        2020 & 6 & 2 & 1 & 0 & 0 \\
        2021 & 4 & 6 & 0 & 0 & 1 \\
        2022 & 5 & 5 & 0 & 0 & 1 \\
        \hline
        Total & 21 & 15 & 2 & 0 & 2 \\
        \bottomrule
    \end{tabular}
    \label{tab:cui_demo}
\end{table}

The objects of study in CUI papers are often (implicitly) WEIRD (Western, Educated, Industrialised, Rich, and Democratic)---where commercially available technologies formed the basis of experiments these were overwhelmingly produced in Europe or North America, and studied for English language tasks. Whilst conducting the literature survey we began to compile a list of global voice assistant launches to demonstrate the diversity of the field (inevitably incomplete, shown in table~\ref{tab:global_vas}); despite the existence of many non-WEIRD CUIs, they remain dramatically unstudied by CUI papers.

\begin{table}
    \centering
    \caption{Global timeline of voice assistant releases.}
    \begin{tabular}{r|l|l}
    \toprule
        Name & Language(s) & Launch \\
        \hline
        Siri & Various (20) & 2011 \\
        Alexa & Various (9) & 2013 \\
        Cortana & Various (9) & 2014 \\
        NUGU & Korean & 2016 \\
        Mycroft & English & 2016 \\
        Clova & Japanese \& Korean & 2017 \\
        Bixby & Various (8) & 2017 \\
        Kakao Mini & Korean & 2017 \\
        Xiao AI & Chinese (Mandarin \& Cantonese) & 2017 \\
        Xiaoyi & Chinese (further details unknown) & 2019 \\
        Celia & Various (4) & 2020 \\
    \bottomrule
    \end{tabular}
    \label{tab:global_vas}
\end{table}

In sum we see that CUI as a conference/community is one that is willing to engage with the ethical aspects of the technologies being studied, but less willing to consider CUI design and evaluation outside of its own geographic context. Provocations are much more likely to address issues of ethics, but these do not seem to be further developed into full papers.

\section{What Can We Do}
There are many forces that shape the design of CUIs and who they serve. Vendors are motivated by profit at the expense of accessibility, data protection regulations designed to protect individuals from organisations indirectly limit multi-user usage, design teams are more aware of issues faced by people like themselves, and the lack of high level perspective results in decisions around e.g. performance of gender that perpetuate existing societal biases. We might not be able to single handedly bring down capitalism, but we \textit{can} alter the research landscape that feeds into industrial research and development and will inspire the next generation of researchers. A key aspect of diversity of geography, culture, gender, ability, and others in the research process is how it shapes \textit{who CUIs are for}. The lack of visibility and consideration for marginalised groups feeds back into who uses and benefits from CUIs, as well as who goes on to research and design them. 

Looking at the causes of the status quo, the situation is understandable from the perspective of an individual researcher. The contextual nature of many of the topics studied by the community (e.g. social interactions and social order) mean that participants must often be drawn from the same culture; widening participation is therefore not simply a matter of balancing participant demographics, studies need to replicated (and potentially also localised) across countries. This conflicts with the pressure to publish---if a paper meets the bar for publication with a single participant group, there is little benefit to replicating it in terms of the metrics by which academics are judged.

Looking back at our own papers when writing this, we found ourselves coming up with excuses. There were prototypes that need to be tested in person, topics that were deeply affected by culture, or language barriers our team simply was not equipped to handle and that local research teams would be. But the reviews by \citeauthor{10.1145/3411764.3445488} and ourselves of CHI/voice assistant publications show us that these arguments apply to at least \textit{three quarters of the community}. We can't rely on authors in non-WEIRD countries to generate the diverse research that we need because those authors \textit{aren't submitting to CUI} (or CHI, or CSCW). Much easier, perhaps, to use provocation papers to point out the inequities that we cannot or will not address in our research (exactly as this paper does)?

At a conference level this begins to more clearly conflict with the stated values of the community. To facilitate attendance from a more diverse group of researchers, CUI offers a 25\% reduction in the conference fee for attendees who live in economically developing countries and allows for remote attendance (CHI similarly offers 35\% and 80\% reductions and virtual attendance). Travel costs, however, are often considerably more than the conference registration, meaning that facilitating virtual attendance is likely to be more effective in terms of diversifying attendance. At the same time, virtual attendance does not present the same quality of conference experience, particularly around informal discussions, networking opportunities, and community building, which risks creating a two tier system that does not address the underlying structural problem. CUI attendees can make use of Gary Marsden Travel Awards to fund in-person travel, but only \textit{once every three years} for physical attendance. More than two thirds of recipients so far from 2023 have been from Europe or North America.

So what else can we do, given that the theme of CUI 2023 is `designing for inclusive conversation'? The surveys suggest that measures focusing on the authors of publications will in turn increase the diversity of participants, as researchers tend to recruit the people around them. Additional financial support for registration and travel for attendees outside WEIRD countries is a key part of this, which could be supported via increased professional registration fees. Other potential options include having parallel submission tracks for more diverse/underrepresented research, as well as to encourage the replication of previously published studies across communities and cultures. This may better align the incentives of researchers, who are pushed to publish, with those of the conference and larger community, who recognise the value of more diverse research and community.

We recognise that these are not prefect answers; to an extent these problems are systemic and a result of the global academic system prioritising North America and Europe (CUI, for example, meets in Europe and uses English as its working language). We are interested to hear suggestions at the conference as to how we might work against this, particularly in ways that align with our other values like sustainability.

\section{Discussion and Reflection}
The discussion and reflections presented here draw from reviews received by the paper, in order to offer a starting point for further discussion at the conference. As per ACM policy we do not directly report the content of reviews, but rather general arguments and suggestions from our reviewers.\footnote{\url{https://www.acm.org/publications/policies/roles-and-responsibilities}}

\subsection{Diversity Is More Than an Institution}
A key issue with trying to `measure' diversity using information available in research papers is that data provided on author affiliations and reported participant demographics is limited and often not directly related to what many would consider to be diversity in this context. Using e.g. the country of an author's institution as a proxy for diversity of culture, background, or ethnicity reduces and shifts the scope of what is implied to be diverse. Analysing participant country of residence similarly glosses over the many different types of diversity that exist within countries, portraying them as homogeneous groups and at a higher level implying that research carried out in a WEIRD context is inherently not diverse. One of the issues at play is the lack of precision around the different kinds of diversity being explored; this can lead to the conflation of apparently similar definitions as well as the reporting of very specific concepts as all-encompassing (e.g. `majority country of residence of the participants on the platform used by a study' as just `diversity of participants').

\subsection{Diversity Is Not a New Problem}
A recurring comment in the reviews was that, as we point out in Section 1, while these issues might not be well represented \textit{at CUI} these kinds of diversity are not a new or unknown problem for HCI and its associated fields of study. There is a rich history of work exploring scientific colonialism and subsequent decolonisation movements that are already present in adjacent conferences like CHI. Indeed, since the original submission of this provocation there has been the CUI@CHI '23 workshop which focused on the inclusive design of CUIs, including submissions relating to global inclusivity~\cite{10.1145/3544549.3573820}. We are delighted to see these discussions beginning to spread from venues where they are already established.

\subsection{Diversity is a Complex Issue}
It was rightly pointed out that this is a very complicated issue with many different components, and that our attempt to offer suggestions in Section 4 smoothed over many of these complexities. Fostering global inclusivity for our research community and those who participate in its research also has political aspects that we did not address: e.g. the structure of the peer-review process inherently makes it harder to publish work that does not conform to WEIRD agendas and methodologies, as the reviewer pool is primarily made up of WEIRD reviewers. Many of the barriers to inclusivity at the level of a conference are often bureaucratic or budgetary. While things may be changing at the top of SIGCHI and ACM, the pace of that change remains very slow. However, this does not mean that we should just wait for change to manifest, and we hope that despite its imperfections this work will be able to provoke discussion and ideas for change at the conference.

\section{Conclusion}
Inspired by work that has mapped out the diversity of the CHI and voice assistant literature, we conducted an informal review of the CUI literature in order to explore how the community approaches issues of ethics and diversity in terms of research, researchers, and participants. While we found a variety of papers covering accessibility, social interaction, and social order, the participants recruited for user studies were overwhelmingly from Europe and North America. We outline contributing factors to the problem of diversity in CUI research and in HCI more broadly, suggesting some starting points for improvement with the aim of seeding a larger discussion at the conference. 

\begin{acks}
This research was funded by the UK Engineering and Physical Sciences Research Council under grant EP/T026723/1.
\end{acks}

\bibliographystyle{ACM-Reference-Format}
\bibliography{main}


\begin{thebibliography}{27}

%%% the Web address.

\ifx \showCODEN    \undefined \def \showCODEN     #1{\unskip}     \fi
\ifx \showDOI      \undefined \def \showDOI       #1{#1}\fi
\ifx \showISBNx    \undefined \def \showISBNx     #1{\unskip}     \fi
\ifx \showISBNxiii \undefined \def \showISBNxiii  #1{\unskip}     \fi
\ifx \showISSN     \undefined \def \showISSN      #1{\unskip}     \fi
\ifx \showLCCN     \undefined \def \showLCCN      #1{\unskip}     \fi
\ifx \shownote     \undefined \def \shownote      #1{#1}          \fi
\ifx \showarticletitle \undefined \def \showarticletitle #1{#1}   \fi
\ifx \showURL      \undefined \def \showURL       {\relax}        \fi
% invisible to TeX
\providecommand\bibfield[2]{#2}
\providecommand\bibinfo[2]{#2}
\providecommand\natexlab[1]{#1}
\providecommand\showeprint[2][]{arXiv:#2}

\bibitem[Abdolrahmani et~al\mbox{.}(2018)]%
        {10.1145/3234695.3236344}
\bibfield{author}{\bibinfo{person}{Ali Abdolrahmani}, \bibinfo{person}{Ravi
  Kuber}, {and} \bibinfo{person}{Stacy~M. Branham}.}
  \bibinfo{year}{2018}\natexlab{}.
\newblock \showarticletitle{"Siri Talks at You": An Empirical Investigation of
  Voice-Activated Personal Assistant (VAPA) Usage by Individuals Who Are
  Blind}. In \bibinfo{booktitle}{\emph{Proceedings of the 20th International
  ACM SIGACCESS Conference on Computers and Accessibility}} (Galway, Ireland)
  \emph{(\bibinfo{series}{ASSETS '18})}. \bibinfo{publisher}{Association for
  Computing Machinery}, \bibinfo{address}{New York, NY, USA},
  \bibinfo{pages}{249–258}.
\newblock
\showISBNx{9781450356503}
\urldef\tempurl%
\url{https://doi.org/10.1145/3234695.3236344}
\showDOI{\tempurl}


\bibitem[Barth et~al\mbox{.}(2020)]%
        {10.1145/3405755.3406128}
\bibfield{author}{\bibinfo{person}{Fabricio Barth}, \bibinfo{person}{Heloisa
  Candello}, \bibinfo{person}{Paulo Cavalin}, {and} \bibinfo{person}{Claudio
  Pinhanez}.} \bibinfo{year}{2020}\natexlab{}.
\newblock \showarticletitle{Intentions, Meanings, and Whys: Designing Content
  for Voice-Based Conversational Museum Guides}. In
  \bibinfo{booktitle}{\emph{Proceedings of the 2nd Conference on Conversational
  User Interfaces}} (Bilbao, Spain) \emph{(\bibinfo{series}{CUI '20})}.
  \bibinfo{publisher}{Association for Computing Machinery},
  \bibinfo{address}{New York, NY, USA}, Article \bibinfo{articleno}{8},
  \bibinfo{numpages}{8}~pages.
\newblock
\showISBNx{9781450375443}
\urldef\tempurl%
\url{https://doi.org/10.1145/3405755.3406128}
\showDOI{\tempurl}


\bibitem[Candello et~al\mbox{.}(2019)]%
        {10.1145/3342775.3342787}
\bibfield{author}{\bibinfo{person}{Heloisa Candello}, \bibinfo{person}{Claudio
  Pinhanez}, \bibinfo{person}{Mauro Pichiliani}, \bibinfo{person}{Marisa
  Vasconcelos}, {and} \bibinfo{person}{Haylla Conde}.}
  \bibinfo{year}{2019}\natexlab{}.
\newblock \showarticletitle{Can Direct Address Affect User Engagement with
  Chatbots Embodied in Physical Spaces?}. In
  \bibinfo{booktitle}{\emph{Proceedings of the 1st International Conference on
  Conversational User Interfaces}} (Dublin, Ireland)
  \emph{(\bibinfo{series}{CUI '19})}. \bibinfo{publisher}{Association for
  Computing Machinery}, \bibinfo{address}{New York, NY, USA}, Article
  \bibinfo{articleno}{3}, \bibinfo{numpages}{9}~pages.
\newblock
\showISBNx{9781450371872}
\urldef\tempurl%
\url{https://doi.org/10.1145/3342775.3342787}
\showDOI{\tempurl}


\bibitem[Catania et~al\mbox{.}(2020)]%
        {10.1145/3405755.3406129}
\bibfield{author}{\bibinfo{person}{Fabio Catania}, \bibinfo{person}{Micol
  Spitale}, \bibinfo{person}{Giulia Cosentino}, {and} \bibinfo{person}{Franca
  Garzotto}.} \bibinfo{year}{2020}\natexlab{}.
\newblock \showarticletitle{What is the Best Action for Children to "Wake Up"
  and "Put to Sleep" a Conversational Agent? A Multi-Criteria Decision Analysis
  Approach}. In \bibinfo{booktitle}{\emph{Proceedings of the 2nd Conference on
  Conversational User Interfaces}} (Bilbao, Spain) \emph{(\bibinfo{series}{CUI
  '20})}. \bibinfo{publisher}{Association for Computing Machinery},
  \bibinfo{address}{New York, NY, USA}, Article \bibinfo{articleno}{4},
  \bibinfo{numpages}{10}~pages.
\newblock
\showISBNx{9781450375443}
\urldef\tempurl%
\url{https://doi.org/10.1145/3405755.3406129}
\showDOI{\tempurl}


\bibitem[Costa and Ribas(2019)]%
        {costa2019ai}
\bibfield{author}{\bibinfo{person}{Pedro Costa} {and}
  \bibinfo{person}{Lu{\'\i}sa Ribas}.} \bibinfo{year}{2019}\natexlab{}.
\newblock \showarticletitle{AI becomes her: Discussing gender and artificial
  intelligence}.
\newblock \bibinfo{journal}{\emph{Technoetic Arts}} \bibinfo{volume}{17},
  \bibinfo{number}{1-2} (\bibinfo{year}{2019}), \bibinfo{pages}{171--193}.
\newblock


\bibitem[Cowan et~al\mbox{.}(2019)]%
        {10.1145/3342775.3342786}
\bibfield{author}{\bibinfo{person}{Benjamin~R. Cowan}, \bibinfo{person}{Philip
  Doyle}, \bibinfo{person}{Justin Edwards}, \bibinfo{person}{Diego Garaialde},
  \bibinfo{person}{Ali Hayes-Brady}, \bibinfo{person}{Holly~P. Branigan},
  \bibinfo{person}{Jo\~{a}o Cabral}, {and} \bibinfo{person}{Leigh Clark}.}
  \bibinfo{year}{2019}\natexlab{}.
\newblock \showarticletitle{What's in an Accent? The Impact of Accented
  Synthetic Speech on Lexical Choice in Human-Machine Dialogue}. In
  \bibinfo{booktitle}{\emph{Proceedings of the 1st International Conference on
  Conversational User Interfaces}} (Dublin, Ireland)
  \emph{(\bibinfo{series}{CUI '19})}. \bibinfo{publisher}{Association for
  Computing Machinery}, \bibinfo{address}{New York, NY, USA}, Article
  \bibinfo{articleno}{23}, \bibinfo{numpages}{8}~pages.
\newblock
\showISBNx{9781450371872}
\urldef\tempurl%
\url{https://doi.org/10.1145/3342775.3342786}
\showDOI{\tempurl}


\bibitem[Du et~al\mbox{.}(2021)]%
        {10.1145/3459990.3465195}
\bibfield{author}{\bibinfo{person}{Yao Du}, \bibinfo{person}{Kerri Zhang},
  \bibinfo{person}{Sruthi Ramabadran}, {and} \bibinfo{person}{Yusa Liu}.}
  \bibinfo{year}{2021}\natexlab{}.
\newblock \showarticletitle{“Alexa, What is That Sound?” A Video Analysis
  of Child-Agent Communication From Two Amazon Alexa Games}
  \emph{(\bibinfo{series}{IDC '21})}. \bibinfo{publisher}{Association for
  Computing Machinery}, \bibinfo{address}{New York, NY, USA},
  \bibinfo{pages}{513–520}.
\newblock
\showISBNx{9781450384520}
\urldef\tempurl%
\url{https://doi.org/10.1145/3459990.3465195}
\showDOI{\tempurl}


\bibitem[Hubbard et~al\mbox{.}(2021)]%
        {10.1145/3469595.3469604}
\bibfield{author}{\bibinfo{person}{Layne Hubbard}, \bibinfo{person}{Shanli
  Ding}, \bibinfo{person}{Vananh Le}, \bibinfo{person}{Pilyoung Kim}, {and}
  \bibinfo{person}{Tom Yeh}.} \bibinfo{year}{2021}\natexlab{}.
\newblock \showarticletitle{Voice Design to Support Young Children’s Agency
  in Child-Agent Interaction}. In \bibinfo{booktitle}{\emph{Proceedings of the
  3rd Conference on Conversational User Interfaces}} (Bilbao (online), Spain)
  \emph{(\bibinfo{series}{CUI '21})}. \bibinfo{publisher}{Association for
  Computing Machinery}, \bibinfo{address}{New York, NY, USA}, Article
  \bibinfo{articleno}{9}, \bibinfo{numpages}{10}~pages.
\newblock
\showISBNx{9781450389983}
\urldef\tempurl%
\url{https://doi.org/10.1145/3469595.3469604}
\showDOI{\tempurl}


\bibitem[Jensen et~al\mbox{.}(2018)]%
        {10.1145/3173574.3173578}
\bibfield{author}{\bibinfo{person}{Rikke~Hagensby Jensen},
  \bibinfo{person}{Yolande Strengers}, \bibinfo{person}{Jesper Kjeldskov},
  \bibinfo{person}{Larissa Nicholls}, {and} \bibinfo{person}{Mikael~B. Skov}.}
  \bibinfo{year}{2018}\natexlab{}.
\newblock \bibinfo{booktitle}{\emph{Designing the Desirable Smart Home: A Study
  of Household Experiences and Energy Consumption Impacts}}.
\newblock \bibinfo{publisher}{Association for Computing Machinery},
  \bibinfo{address}{New York, NY, USA}, \bibinfo{pages}{1--14}.
\newblock
\showISBNx{9781450356206}
\urldef\tempurl%
\url{https://doi.org/10.1145/3173574.3173578}
\showURL{%
\tempurl}


\bibitem[Jestin et~al\mbox{.}(2022)]%
        {10.1145/3543829.3543836}
\bibfield{author}{\bibinfo{person}{Iris Jestin}, \bibinfo{person}{Joel
  Fischer}, \bibinfo{person}{Maria~Jose Galvez~Trigo}, \bibinfo{person}{David
  Large}, {and} \bibinfo{person}{Gary Burnett}.}
  \bibinfo{year}{2022}\natexlab{}.
\newblock \showarticletitle{Effects of Wording and Gendered Voices on
  Acceptability of Voice Assistants in Future Autonomous Vehicles}. In
  \bibinfo{booktitle}{\emph{Proceedings of the 4th Conference on Conversational
  User Interfaces}} (Glasgow, United Kingdom) \emph{(\bibinfo{series}{CUI
  '22})}. \bibinfo{publisher}{Association for Computing Machinery},
  \bibinfo{address}{New York, NY, USA}, Article \bibinfo{articleno}{24},
  \bibinfo{numpages}{11}~pages.
\newblock
\showISBNx{9781450397391}
\urldef\tempurl%
\url{https://doi.org/10.1145/3543829.3543836}
\showDOI{\tempurl}


\bibitem[Lee(2020)]%
        {10.1145/3405755.3406124}
\bibfield{author}{\bibinfo{person}{Minha Lee}.}
  \bibinfo{year}{2020}\natexlab{}.
\newblock \showarticletitle{Speech Acts Redux: Beyond Request-Response
  Interactions}. In \bibinfo{booktitle}{\emph{Proceedings of the 2nd Conference
  on Conversational User Interfaces}} (Bilbao, Spain)
  \emph{(\bibinfo{series}{CUI '20})}. \bibinfo{publisher}{Association for
  Computing Machinery}, \bibinfo{address}{New York, NY, USA}, Article
  \bibinfo{articleno}{13}, \bibinfo{numpages}{10}~pages.
\newblock
\showISBNx{9781450375443}
\urldef\tempurl%
\url{https://doi.org/10.1145/3405755.3406124}
\showDOI{\tempurl}


\bibitem[Lee et~al\mbox{.}(2022)]%
        {10.1145/3543829.3543835}
\bibfield{author}{\bibinfo{person}{Minha Lee}, \bibinfo{person}{Lily Frank},
  \bibinfo{person}{Yvonne De~Kort}, {and} \bibinfo{person}{Wijnand
  IJsselsteijn}.} \bibinfo{year}{2022}\natexlab{}.
\newblock \showarticletitle{Where is Vincent? Expanding Our Emotional Selves
  with AI}. In \bibinfo{booktitle}{\emph{Proceedings of the 4th Conference on
  Conversational User Interfaces}} (Glasgow, United Kingdom)
  \emph{(\bibinfo{series}{CUI '22})}. \bibinfo{publisher}{Association for
  Computing Machinery}, \bibinfo{address}{New York, NY, USA}, Article
  \bibinfo{articleno}{19}, \bibinfo{numpages}{11}~pages.
\newblock
\showISBNx{9781450397391}
\urldef\tempurl%
\url{https://doi.org/10.1145/3543829.3543835}
\showDOI{\tempurl}


\bibitem[Lima et~al\mbox{.}(2019)]%
        {10.1145/3308560.3317597}
\bibfield{author}{\bibinfo{person}{Lanna Lima}, \bibinfo{person}{Vasco
  Furtado}, \bibinfo{person}{Elizabeth Furtado}, {and}
  \bibinfo{person}{Virgilio Almeida}.} \bibinfo{year}{2019}\natexlab{}.
\newblock \showarticletitle{Empirical Analysis of Bias in Voice-Based Personal
  Assistants}. In \bibinfo{booktitle}{\emph{Companion Proceedings of The 2019
  World Wide Web Conference}} (San Francisco, USA) \emph{(\bibinfo{series}{WWW
  '19})}. \bibinfo{publisher}{Association for Computing Machinery},
  \bibinfo{address}{New York, NY, USA}, \bibinfo{pages}{533–538}.
\newblock
\showISBNx{9781450366755}
\urldef\tempurl%
\url{https://doi.org/10.1145/3308560.3317597}
\showDOI{\tempurl}


\bibitem[Linxen et~al\mbox{.}(2021)]%
        {10.1145/3411764.3445488}
\bibfield{author}{\bibinfo{person}{Sebastian Linxen},
  \bibinfo{person}{Christian Sturm}, \bibinfo{person}{Florian Br\"{u}hlmann},
  \bibinfo{person}{Vincent Cassau}, \bibinfo{person}{Klaus Opwis}, {and}
  \bibinfo{person}{Katharina Reinecke}.} \bibinfo{year}{2021}\natexlab{}.
\newblock \bibinfo{booktitle}{\emph{How WEIRD is CHI?}}
\newblock \bibinfo{publisher}{Association for Computing Machinery},
  \bibinfo{address}{New York, NY, USA}.
\newblock
\showISBNx{9781450380966}
\urldef\tempurl%
\url{https://doi.org/10.1145/3411764.3445488}
\showURL{%
\tempurl}


\bibitem[Lovato and Piper(2015)]%
        {10.1145/2771839.2771910}
\bibfield{author}{\bibinfo{person}{Silvia Lovato} {and}
  \bibinfo{person}{Anne~Marie Piper}.} \bibinfo{year}{2015}\natexlab{}.
\newblock \showarticletitle{"Siri, is This You?": Understanding Young
  Children's Interactions with Voice Input Systems}. In
  \bibinfo{booktitle}{\emph{Proceedings of the 14th International Conference on
  Interaction Design and Children}} (Boston, Massachusetts)
  \emph{(\bibinfo{series}{IDC '15})}. \bibinfo{publisher}{Association for
  Computing Machinery}, \bibinfo{address}{New York, NY, USA},
  \bibinfo{pages}{335–338}.
\newblock
\showISBNx{9781450335904}
\urldef\tempurl%
\url{https://doi.org/10.1145/2771839.2771910}
\showDOI{\tempurl}


\bibitem[McKay and Miller(2021)]%
        {10.1145/3411764.3445114}
\bibfield{author}{\bibinfo{person}{Dana McKay} {and} \bibinfo{person}{Charlynn
  Miller}.} \bibinfo{year}{2021}\natexlab{}.
\newblock \showarticletitle{Standing in the Way of Control: A Call to Action to
  Prevent Abuse through Better Design of Smart Technologies}. In
  \bibinfo{booktitle}{\emph{Proceedings of the 2021 CHI Conference on Human
  Factors in Computing Systems}} (Yokohama, Japan) \emph{(\bibinfo{series}{CHI
  '21})}. \bibinfo{publisher}{Association for Computing Machinery},
  \bibinfo{address}{New York, NY, USA}, Article \bibinfo{articleno}{332},
  \bibinfo{numpages}{14}~pages.
\newblock
\showISBNx{9781450380966}
\urldef\tempurl%
\url{https://doi.org/10.1145/3411764.3445114}
\showDOI{\tempurl}


\bibitem[Nass et~al\mbox{.}(1994)]%
        {nass1994computers}
\bibfield{author}{\bibinfo{person}{Clifford Nass}, \bibinfo{person}{Jonathan
  Steuer}, {and} \bibinfo{person}{Ellen~R. Tauber}.}
  \bibinfo{year}{1994}\natexlab{}.
\newblock \showarticletitle{Computers Are Social Actors}. In
  \bibinfo{booktitle}{\emph{Proceedings of the SIGCHI Conference on Human
  Factors in Computing Systems}} (Boston, Massachusetts, USA)
  \emph{(\bibinfo{series}{CHI '94})}. \bibinfo{publisher}{Association for
  Computing Machinery}, \bibinfo{address}{New York, NY, USA},
  \bibinfo{pages}{72–78}.
\newblock
\showISBNx{0897916506}
\urldef\tempurl%
\url{https://doi.org/10.1145/191666.191703}
\showDOI{\tempurl}


\bibitem[Pradhan et~al\mbox{.}(2018)]%
        {10.1145/3173574.3174033}
\bibfield{author}{\bibinfo{person}{Alisha Pradhan}, \bibinfo{person}{Kanika
  Mehta}, {and} \bibinfo{person}{Leah Findlater}.}
  \bibinfo{year}{2018}\natexlab{}.
\newblock \bibinfo{booktitle}{\emph{"Accessibility Came by Accident": Use of
  Voice-Controlled Intelligent Personal Assistants by People with
  Disabilities}}.
\newblock \bibinfo{publisher}{Association for Computing Machinery},
  \bibinfo{address}{New York, NY, USA}, \bibinfo{pages}{1–13}.
\newblock
\showISBNx{9781450356206}
\urldef\tempurl%
\url{https://doi.org/10.1145/3173574.3174033}
\showURL{%
\tempurl}


\bibitem[Rinc\'{o}n et~al\mbox{.}(2021)]%
        {10.1145/3449206}
\bibfield{author}{\bibinfo{person}{Cami Rinc\'{o}n}, \bibinfo{person}{Os
  Keyes}, {and} \bibinfo{person}{Corinne Cath}.}
  \bibinfo{year}{2021}\natexlab{}.
\newblock \showarticletitle{Speaking from Experience: Trans/Non-Binary
  Requirements for Voice-Activated AI}.
\newblock \bibinfo{journal}{\emph{Proc. ACM Hum.-Comput. Interact.}}
  \bibinfo{volume}{5}, \bibinfo{number}{CSCW1}, Article
  \bibinfo{articleno}{132} (\bibinfo{date}{apr} \bibinfo{year}{2021}),
  \bibinfo{numpages}{27}~pages.
\newblock
\urldef\tempurl%
\url{https://doi.org/10.1145/3449206}
\showDOI{\tempurl}


\bibitem[Sayago and Ribera(2020)]%
        {10.1145/3439231.3440603}
\bibfield{author}{\bibinfo{person}{Sergio Sayago} {and} \bibinfo{person}{Mireia
  Ribera}.} \bibinfo{year}{2020}\natexlab{}.
\newblock \showarticletitle{Apple Siri (Input) + Voice Over (Output) = a de
  Facto Marriage: An Exploratory Case Study with Blind People}. In
  \bibinfo{booktitle}{\emph{9th International Conference on Software
  Development and Technologies for Enhancing Accessibility and Fighting
  Info-Exclusion}} (Online, Portugal) \emph{(\bibinfo{series}{DSAI 2020})}.
  \bibinfo{publisher}{Association for Computing Machinery},
  \bibinfo{address}{New York, NY, USA}, \bibinfo{pages}{6–10}.
\newblock
\showISBNx{9781450389372}
\urldef\tempurl%
\url{https://doi.org/10.1145/3439231.3440603}
\showDOI{\tempurl}


\bibitem[Schneider(2021)]%
        {schneider2021recommended}
\bibfield{author}{\bibinfo{person}{Florian Schneider}.}
  \bibinfo{year}{2021}\natexlab{}.
\newblock \showarticletitle{Recommended by Google Home}. In
  \bibinfo{booktitle}{\emph{International Conference on Human-Computer
  Interaction}}. Springer, \bibinfo{pages}{485--493}.
\newblock


\bibitem[Sciuto et~al\mbox{.}(2018)]%
        {10.1145/3196709.3196772}
\bibfield{author}{\bibinfo{person}{Alex Sciuto}, \bibinfo{person}{Arnita
  Saini}, \bibinfo{person}{Jodi Forlizzi}, {and} \bibinfo{person}{Jason~I.
  Hong}.} \bibinfo{year}{2018}\natexlab{}.
\newblock \showarticletitle{"Hey Alexa, What's Up?": A Mixed-Methods Studies of
  In-Home Conversational Agent Usage}. In \bibinfo{booktitle}{\emph{Proceedings
  of the 2018 Designing Interactive Systems Conference}} (Hong Kong, China)
  \emph{(\bibinfo{series}{DIS '18})}. \bibinfo{publisher}{Association for
  Computing Machinery}, \bibinfo{address}{New York, NY, USA},
  \bibinfo{pages}{857–868}.
\newblock
\showISBNx{9781450351980}
\urldef\tempurl%
\url{https://doi.org/10.1145/3196709.3196772}
\showDOI{\tempurl}


\bibitem[Seymour et~al\mbox{.}(2023)]%
        {seymour2022systematic}
\bibfield{author}{\bibinfo{person}{William Seymour}, \bibinfo{person}{Xiao
  Zhan}, \bibinfo{person}{Mark Cote}, {and} \bibinfo{person}{Jose Such}.}
  \bibinfo{year}{2023}\natexlab{}.
\newblock \showarticletitle{A Systematic Review of Ethical Concerns with Voice
  Assistants}. In \bibinfo{booktitle}{\emph{Accepted to the 2023 AAAI/ACM
  Conference on Artificial Intelligence, Ethics, and Society}}.
\newblock
\urldef\tempurl%
\url{https://arxiv.org/abs/2211.04193}
\showURL{%
\tempurl}


\bibitem[Sin et~al\mbox{.}(2023)]%
        {10.1145/3544549.3573820}
\bibfield{author}{\bibinfo{person}{Jaisie Sin}, \bibinfo{person}{Heloisa
  Candello}, \bibinfo{person}{Leigh Clark}, \bibinfo{person}{Benjamin~R.
  Cowan}, \bibinfo{person}{Minha Lee}, \bibinfo{person}{Cosmin Munteanu},
  \bibinfo{person}{Martin Porcheron}, \bibinfo{person}{Sarah~Theres
  V\"{o}lkel}, \bibinfo{person}{Stacy Branham}, \bibinfo{person}{Robin~N.
  Brewer}, \bibinfo{person}{Ana~Paula Chaves}, \bibinfo{person}{Razan Jaber},
  {and} \bibinfo{person}{Amanda Lazar}.} \bibinfo{year}{2023}\natexlab{}.
\newblock \showarticletitle{CUI@CHI: Inclusive Design of CUIs Across Modalities
  and Mobilities}. In \bibinfo{booktitle}{\emph{Extended Abstracts of the 2023
  CHI Conference on Human Factors in Computing Systems}} (Hamburg, Germany)
  \emph{(\bibinfo{series}{CHI EA '23})}. \bibinfo{publisher}{Association for
  Computing Machinery}, \bibinfo{address}{New York, NY, USA}, Article
  \bibinfo{articleno}{341}, \bibinfo{numpages}{5}~pages.
\newblock
\showISBNx{9781450394222}
\urldef\tempurl%
\url{https://doi.org/10.1145/3544549.3573820}
\showDOI{\tempurl}


\bibitem[Sutton(2020)]%
        {10.1145/3405755.3406123}
\bibfield{author}{\bibinfo{person}{Selina~Jeanne Sutton}.}
  \bibinfo{year}{2020}\natexlab{}.
\newblock \showarticletitle{Gender Ambiguous, Not Genderless: Designing Gender
  in Voice User Interfaces (VUIs) with Sensitivity}. In
  \bibinfo{booktitle}{\emph{Proceedings of the 2nd Conference on Conversational
  User Interfaces}} (Bilbao, Spain) \emph{(\bibinfo{series}{CUI '20})}.
  \bibinfo{publisher}{Association for Computing Machinery},
  \bibinfo{address}{New York, NY, USA}, Article \bibinfo{articleno}{11},
  \bibinfo{numpages}{8}~pages.
\newblock
\showISBNx{9781450375443}
\urldef\tempurl%
\url{https://doi.org/10.1145/3405755.3406123}
\showDOI{\tempurl}


\bibitem[Weizenbaum(1966)]%
        {weizenbaum1966eliza}
\bibfield{author}{\bibinfo{person}{Joseph Weizenbaum}.}
  \bibinfo{year}{1966}\natexlab{}.
\newblock \showarticletitle{ELIZA-a Computer Program for the Study of Natural
  Language Communication between Man and Machine}.
\newblock \bibinfo{journal}{\emph{Commun. ACM}} \bibinfo{volume}{9},
  \bibinfo{number}{1} (\bibinfo{date}{Jan.} \bibinfo{year}{1966}),
  \bibinfo{pages}{36--45}.
\newblock
\showISSN{0001-0782}
\urldef\tempurl%
\url{https://doi.org/10.1145/365153.365168}
\showDOI{\tempurl}


\bibitem[Wu et~al\mbox{.}(2022)]%
        {10.1145/3543829.3543839}
\bibfield{author}{\bibinfo{person}{Yunhan Wu}, \bibinfo{person}{Martin
  Porcheron}, \bibinfo{person}{Philip Doyle}, \bibinfo{person}{Justin Edwards},
  \bibinfo{person}{Daniel Rough}, \bibinfo{person}{Orla Cooney},
  \bibinfo{person}{Anna Bleakley}, \bibinfo{person}{Leigh Clark}, {and}
  \bibinfo{person}{Benjamin Cowan}.} \bibinfo{year}{2022}\natexlab{}.
\newblock \showarticletitle{Comparing Command Construction in Native and
  Non-Native Speaker IPA Interaction through Conversation Analysis}. In
  \bibinfo{booktitle}{\emph{Proceedings of the 4th Conference on Conversational
  User Interfaces}} (Glasgow, United Kingdom) \emph{(\bibinfo{series}{CUI
  '22})}. \bibinfo{publisher}{Association for Computing Machinery},
  \bibinfo{address}{New York, NY, USA}, Article \bibinfo{articleno}{10},
  \bibinfo{numpages}{12}~pages.
\newblock
\showISBNx{9781450397391}
\urldef\tempurl%
\url{https://doi.org/10.1145/3543829.3543839}
\showDOI{\tempurl}


\end{thebibliography}

\end{document}